\documentclass[%
 reprint,
 amsmath,amssymb,
 aps,
]{revtex4-2}
\usepackage{times}
\usepackage{graphicx}% Include figure files
\usepackage{dcolumn}% Align table columns on decimal point
\usepackage{bm}% bold math
\usepackage{orcidlink}
\usepackage{xcolor}
\usepackage{braket}
\usepackage{hyperref}
\usepackage{mathtools}
\allowdisplaybreaks
\begin{document}

\preprint{APS/123-QED}

\title{Perturbative Input–Output Theory of Floquet Cavity Magnonics and Magnon Energy Shifts}% Force line breaks with \\

\author{T. Aguiar %
 \orcidlink{0000-0002-1141-2241}}
\email{taguiarcf@gmail.com}
\affiliation{Instituto de F\'\i sica  Gleb Wataghin, Universidade Estadual de Campinas, 13083-859, Campinas, SP, Brazil} %\altaffiliation[Also at ]{Physics Department, XYZ University.}%Lines break automatically or can be forced with \\
\author{M. C. de Oliveira%
 \orcidlink{0000-0003-2251-2632}}\email{marcos@ifi.unicamp.br}
\affiliation{Instituto de F\'\i sica  Gleb Wataghin, Universidade Estadual de Campinas, 13083-859, Campinas, SP, Brazil}%

\date{\today}% It is always \today, today,
\begin{abstract}
We develop a perturbative input--output formalism to compute the reflectance and transmittance spectra of cavity magnonics systems subject to a Floquet modulation. The method exploits the strong hierarchy between the magnetic-dipole couplings transverse (drive field) and parallel (modulation field) to the static bias field, which naturally introduces the small parameter $\epsilon = (2Ns)^{-1/2}$ associated with the total spin $Ns$ of the ferromagnet. By organizing the cavity and magnon fields in a systematic expansion in $\epsilon$, we obtain compact analytic expressions for the spectra up to second order. Using these results, we reproduce the characteristic sideband structure observed in recent Floquet cavity electromagnonics experiments \cite{xu2020floquet}. Furthermore, accounting for the Zeeman interaction between the modulation field and the fully polarized ground state---a contribution typically neglected in previous treatments—we predict an additional magnon detuning of approximately $0.8\,\mathrm{GHz}$, independent of both modulation frequency and sample size and determined solely by the spatial volume occupied by the modulation field and the saturation magnetization of the material sample $M_s$. This identifies a measurable and previously overlooked shift relevant for the interpretation and design of cavity magnonics experiments.
\end{abstract}

%Use showkeys class option if keyword
\maketitle

%\tableofcontents

\section{\label{sec:level1} Introduction}

The field of quantum magnonics has recently advanced into a regime where coherent hybrid magnon--photon states can be prepared, controlled, and probed under Floquet driving~\cite{xu2020floquet,xu2023dynamical,pishehvar2025resonance,pishehvar2025demand,zhuang2024dynamical}. 
In these experiments, a ferromagnet subjected to a static bias field is periodically driven by an additional oscillatory magnetic field aligned with the bias direction, resulting in a time-periodic modulation of its magnetization dynamics. 
Such Floquet modulation enables the observation of rich physical phenomena—including the emergence of sidebands, Rabi oscillations \cite{Rabi_osc_mag}, and Autler--Townes splittings—that closely mirror effects familiar from atomic and molecular physics \cite{towner1955microwave,AT_revisited,agarwal1985vacuum}.

\begin{figure}[ht]
    \centering
    \includegraphics[width=0.8\linewidth]{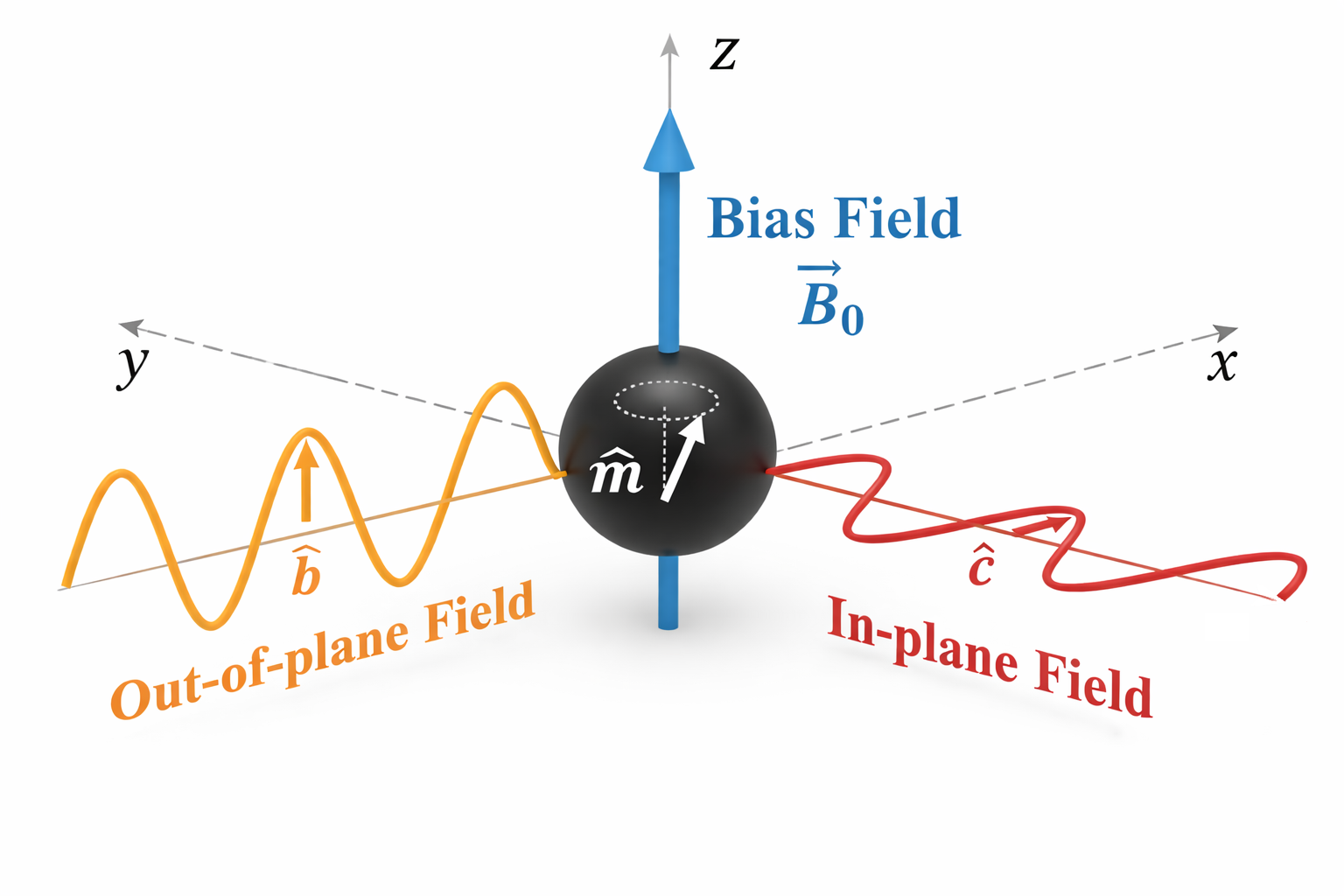}
    \caption{Illustration of the ferromagnet excited by the two microwave fields.}
    \label{fig:schematic}
\end{figure}

The physical setting underlying Floquet electromagnonics typically consists of a ferromagnetic sphere interacting with three electromagnetic fields:  
(i) a strong static field that polarizes the magnet and sets the natural precession frequency;  
(ii) an in-plane microwave field, perpendicular to the bias, that drives coherent magnon excitation; and  
(iii) an out-of-plane (parallel to the bias) modulation field that periodically perturbs the precession frequency (See Fig.\ref{fig:schematic}).
Through the combined action of these three inputs, cavity magnonics experiments have achieved remarkable levels of quantum control, including the resolution of single spin flips in macroscopic magnets~\cite{lachance2017resolving,lachance2020entanglement} and the manipulation of coherent superpositions of distinct magnon modes~\cite{xu2023quantum}. 
These developments position cavity magnonics as a promising platform for hybrid quantum information processing \cite{chumak2014magnon,chumak2022advances} and for exploring driven light--matter interactions in solid-state systems.

In this work, we develop an experimentally motivated perturbative framework for modeling the reflectance and transmittance spectra of cavity magnonics systems under Floquet driving. 
By rewriting the magnon--photon Hamiltonian in terms of spin-enhanced couplings to the uniform magnetostatic mode, we identify a natural small parameter, $\epsilon=(2Ns)^{-1/2}$, where $N$ is the number of atomic sites and $s$ the spin per site.  
This enables a compact and physically transparent expansion of the cavity and magnon fields in well-organized orders of~$\epsilon$, allowing us to derive analytic expressions for the spectral response and to expose contributions that are typically obscured in fully numerical treatments.

Beyond providing a compact description of Floquet-driven spectra, the perturbative approach developed here is broadly applicable to a wide class of cavity magnonics platforms. Because the method organizes the dynamics in powers of the spin-enhanced couplings, it naturally scales to systems with multiple cavity modes, several magnon branches, or additional quantum degrees of freedom. This makes it especially relevant for hybrid quantum technologies in which magnons act as coherent transducers, memories, or nonlinear elements. In such architectures, an analytic framework that captures Floquet engineering, multi-photon processes, and frequency shifts is essential for designing and optimizing future quantum devices.

The structure of the paper is as follows.  
In Sec.~\ref{hierchy}, we establish the hierarchy of spin-enhanced couplings by deriving the effective magnon Hamiltonian in the presence of in-plane and out-of-plane oscillatory fields.  
In Sec.~\ref{pert_ref}, we construct the perturbative input–output formalism and obtain the spectral equations governing the cavity and magnon fields.  
In Sec.~\ref{discussion}, we apply the method to calculate second-order reflectance spectra and compare our results with experimental observations.  
Finally, Sec.~\ref{conc} summarizes the main findings and discusses implications for future cavity magnonics experiments.

\section{Hierarchy between the Spin-Enhanced Couplings}\label{hierchy}

In microwave cavity magnonics experiments, a ferromagnet is positioned at a point of maximal microwave magnetic field so that the induced spin excitations are nearly uniform across the sample and can be described by plane-wave modes. 
We consider a ferromagnet fully polarized along the $z$ axis and interacting with two quantized electromagnetic fields: an in-plane field (drive) and an out-of-plane field (Floquet modulation). 
The microscopic Hamiltonian is
\begin{align}
\hat{H} = 
&-J \sum_{\boldsymbol{\mathfrak{i}},\boldsymbol{\mathfrak{j}}} 
\hat{\boldsymbol{S}}_{\boldsymbol{\mathfrak{i}}}\!\cdot\!
\hat{\boldsymbol{S}}_{\boldsymbol{\mathfrak{i}}+\boldsymbol{\mathfrak{j}}}
- g_0 \sum_{\boldsymbol{\mathfrak{i}}} \hat{S}^z_{\boldsymbol{\mathfrak{i}}} \nonumber\\
&- g_c \sum_{\boldsymbol{\mathfrak{i}}} \hat{S}^x_{\boldsymbol{\mathfrak{i}}}
\!\left( \hat{c}_{\boldsymbol{k_c}} e^{i\boldsymbol{k_c}\cdot \boldsymbol{r_{\mathfrak{i}}}}
+ \hat{c}^\dagger_{\boldsymbol{k_c}} e^{-i\boldsymbol{k_c}\cdot \boldsymbol{r_{\mathfrak{i}}}} \right) \nonumber\\
&- g_b \sum_{\boldsymbol{\mathfrak{i}}} \hat{S}^z_{\boldsymbol{\mathfrak{i}}}
\!\left( \hat{b}_{\boldsymbol{k_b}} e^{i\boldsymbol{k_b}\cdot \boldsymbol{r_{\mathfrak{i}}}}
+ \hat{b}^\dagger_{\boldsymbol{k_b}} e^{-i\boldsymbol{k_b}\cdot \boldsymbol{r_{\mathfrak{i}}}} \right),
\end{align}
where $\boldsymbol{r_{\mathfrak{i}}}$ denotes the position of the $\mathfrak{i}$-th lattice site and $\boldsymbol{\mathfrak{j}}$ runs over its nearest neighbors.  
The operators $\hat{c}$ and $\hat{b}$ annihilate photons of the in-plane and out-of-plane fields, respectively; $g_c$ and $g_b$ are the corresponding intrinsic magnetic-dipole couplings, and $g_0=\gamma B_0$ is the static Zeeman coupling.  

The usual wavelengths in microwave cavity setups are in the $\text{GHz}$ range, whereby the wavelengths are in the order of $\lambda_c \sim 300\text{ mm}$.  For this article, the YIG sphere that houses the uniform magnon mode has a diameter in the order of $400~\mu\text{m}=0.4\text{ mm}\ll \lambda_c$ . This justifies the assumption that the magnetic field within the ferrimagnet is approximately uniform.  Therefore, we approximate the field operators by their spatially uniform form and write
\begin{align}
\hat{H} \approx 
&-J \sum_{\boldsymbol{\mathfrak{i}},\boldsymbol{\mathfrak{j}}} 
\hat{\boldsymbol{S}}_{\boldsymbol{\mathfrak{i}}}\!\cdot\!
\hat{\boldsymbol{S}}_{\boldsymbol{\mathfrak{i}}+\boldsymbol{\mathfrak{j}}}
- g_0 \sum_{\boldsymbol{\mathfrak{i}}} \hat{S}^z_{\boldsymbol{\mathfrak{i}}} \nonumber\\
&- g_c\sum_{\boldsymbol{\mathfrak{i}}}\hat{S}^x_{\boldsymbol{\mathfrak{i}}}(\hat{c}+\hat{c}^\dagger)
- g_b\sum_{\boldsymbol{\mathfrak{i}}}\hat{S}^z_{\boldsymbol{\mathfrak{i}}}(\hat{b}+\hat{b}^\dagger).
\end{align}

To express the Hamiltonian in terms of magnon operators, we re-define the $x$ and $y$ spin components in terms of ladder operators, $\hat{S}_{\boldsymbol{\mathfrak{i}}}^x=(\hat{S}_{\boldsymbol{\mathfrak{i}}}^++\hat{S}_{\boldsymbol{\mathfrak{i}}}^-)/2$ ,  $\hat{S}_{\boldsymbol{\mathfrak{i}}}^y=(\hat{S}_{\boldsymbol{\mathfrak{i}}}^+-\hat{S}_{\boldsymbol{\mathfrak{i}}}^-)/2i$, and apply the Holstein--Primakoff (HP) transformation
\begin{align}
\hat{S}^+_{\boldsymbol{\mathfrak{i}}}
&\approx (2s)^{1/2}\!\left( \mathbb{I} - \frac{\hat{m}^\dagger_{\boldsymbol{\mathfrak{i}}}
\hat{m}_{\boldsymbol{\mathfrak{i}}}}{4s} \right) \hat{m}_{\boldsymbol{\mathfrak{i}}},\\
\hat{S}^-_{\boldsymbol{\mathfrak{i}}}
&\approx (2s)^{1/2}\hat{m}^\dagger_{\boldsymbol{\mathfrak{i}}}
\!\left( \mathbb{I} - \frac{\hat{m}^\dagger_{\boldsymbol{\mathfrak{i}}}
\hat{m}_{\boldsymbol{\mathfrak{i}}}}{4s} \right),\\
\hat{S}^z_{\boldsymbol{\mathfrak{i}}}
&\rightarrow s - \hat{m}^\dagger_{\boldsymbol{\mathfrak{i}}}\hat{m}_{\boldsymbol{\mathfrak{i}}},
\end{align}
which maps spin operators at each site onto bosonic spin-flip operators $\hat{m}_{\boldsymbol{\mathfrak{i}}}$ and $\hat{m}^\dagger_{\boldsymbol{\mathfrak{i}}}$.  
We retain the second-order term in the HP expansion because it contributes to the second-order reflectance calculated in Sec.~\ref{sec_ord}.

Switching to momentum space and focusing on the uniform magnetostatic mode, we obtain the effective single-magnon-mode Hamiltonian
\begin{align}
\hat{H} =
&\, \omega_m \hat{m}^\dagger \hat{m}
- g_0 Ns\,\mathbb{I}
- g_b Ns(\hat{b}+\hat{b}^\dagger) \nonumber\\
&+ g_b \hat{m}^\dagger \hat{m} (\hat{b}+\hat{b}^\dagger)
- \tfrac{1}{2} g_c (2Ns)^{1/2}
(\hat{m}+\hat{m}^\dagger)(\hat{c}+\hat{c}^\dagger) \nonumber\\
&+ \frac{g_c}{4(2Ns)^{1/2}}
\bigl(\hat{m}^\dagger \hat{m}^2 + \hat{m}^{\dagger 2}\hat{m}\bigr)
(\hat{c}+\hat{c}^\dagger).
\end{align}
It is convenient to introduce the perturbative parameter
$\epsilon = (2Ns)^{-1/2}$  
and the spin-enhanced couplings
\begin{align}
G_b = (2Ns)^{1/2} g_b = \epsilon^{-1} g_b,\qquad
G_c = \frac{g_c}{2}\epsilon^{-1},
\end{align}
where
\begin{align}
    g_{b,c}=\gamma\sqrt{\dfrac{\mu_0\hbar \omega_{b,c}}{2V_{b,c}}},
    \label{g_def}
\end{align}
such that $\gamma=28.02\text{ GHz}/\text{T}$ is the electron gyromagnetic ratio, $\mu_0=4\pi \times 10^{-7}$ is the vacuum magnetic permeability, $\omega_{b,c}$/$V_{b,c}$ are the frequency/effective volume respective to the out-of-plane ($b$) and in-plane field fields, respectively. Since intrinsic magnetic-dipole couplings are typically of order $10~\mathrm{mHz}$~\cite{tabuchi2016quantum} and $N\sim 10^{18}$ for millimeter-scale YIG spheres~\cite{xu2020floquet}, both $G_b$ and $G_c$ lie in the $\text{MHz}$ range.  
Expressed in these variables, the Hamiltonian becomes
\begin{align}
\hat{H} =
&\, \omega_m \hat{m}^\dagger\hat{m}
- g_0 Ns\, \mathbb{I}
- \tfrac{1}{2}\epsilon^{-1} G_b (\hat{b}+\hat{b}^\dagger) \nonumber\\
&+ \epsilon G_b \hat{m}^\dagger\hat{m}(\hat{b}+\hat{b}^\dagger)
- G_c (\hat{m}+\hat{m}^\dagger)(\hat{c}+\hat{c}^\dagger) \nonumber\\
&+ \frac{G_c}{4Ns}
\left( \hat{m}^\dagger\hat{m}^2 + \hat{m}^{\dagger 2}\hat{m} \right)
(\hat{c}+\hat{c}^\dagger).
\end{align}

The $\epsilon^{-1}$ term corresponds to the Zeeman interaction between the modulation field and the fully polarized ground state of the ferromagnet—a contribution typically omitted in cavity magnonics treatments.  
As we show later, this term leads to a measurable detuning of the magnon resonance.

When the in-plane field is tuned near the magnon frequency, we apply the rotating-wave approximation (RWA) to isolate the resonant processes.  The RWA Hamiltonian becomes
\begin{align}
\hat{H}^{(\mathrm{RWA})} =
&\, \hat{H}_0^{(\mathrm{RWA})}
- \tfrac{1}{2}\epsilon^{-1} G_b (\hat{b}+\hat{b}^\dagger) \nonumber\\
&+ \epsilon G_b (\hat{b}+\hat{b}^\dagger)\hat{m}^\dagger\hat{m} \nonumber\\
&+ \tfrac{1}{2}\epsilon^2 G_c
\left( \hat{m}^{\dagger 2}\hat{m}\,\hat{c}
+ \hat{m}^\dagger\hat{m}^2\,\hat{c}^\dagger \right),
\label{RWA_ham}
\end{align}
where
\begin{align}
\hat{H}_0^{(\mathrm{RWA})} =
\omega_c \hat{c}^\dagger\hat{c}
+ \omega_m \hat{m}^\dagger\hat{m}
+ \omega_b \hat{b}^\dagger\hat{b}
- G_c(\hat{m}\hat{c}^\dagger+\hat{m}^\dagger\hat{c}).
\end{align}
The hierarchy of terms in powers of $\epsilon$ will form the basis of the perturbative input--output expansion developed in Sec.~\ref{pert_ref}.

\section{Perturbative Input--Output Formalism\label{pert_ref}}

The most direct application of the theory developed in this work is to provide an alternative, analytically tractable method to model the reflectance and transmittance of cavity magnonics experiments.  
To this end, we employ the input--output formalism of a two-port cavity~\cite{milburn2012quantum} and write the Heisenberg--Langevin equations of motion for the in-plane, magnon, and out-of-plane (modulation) fields, denoted by the operators $\hat{c}$, $\hat{m}$, and $\hat{b}$, respectively.  
Collecting them into the vector
\(
(\hat{c},\hat{m},\hat{b})^{\mathsf{T}},
\)
we obtain
\begin{align}
\left( \frac{d}{dt} + \hat{A} \right)
\begin{bmatrix}
\hat{c} \\
\hat{m} \\
\hat{b}
\end{bmatrix}
=&\,
\hat{A}_{\text{in}}
+ \frac{i}{2} G_b \epsilon^{-1}
\begin{bmatrix}
0\\[2pt]
0\\[2pt]
1
\end{bmatrix} \mathbb{I}
- i G_c
\begin{bmatrix}
\hat{m}\\
\hat{c}\\
0
\end{bmatrix} \nonumber\\[4pt]
&- i G_b \epsilon
\begin{bmatrix}
0\\[2pt]
(\hat{b}+\hat{b}^\dagger)\hat{m}\\[2pt]
\hat{m}^\dagger \hat{m}
\end{bmatrix}\nonumber\\
&- \frac{i}{2}\epsilon^2 G_c
\begin{bmatrix}
\hat{m}^\dagger \hat{m}^2\\[2pt]
2\hat{m}^\dagger \hat{m}\hat{c} + \hat{m}^2 \hat{c}^\dagger\\[2pt]
0
\end{bmatrix},
\end{align}
where
\begin{align}
\hat{A}_{\text{in}} \equiv
\begin{bmatrix}
\sqrt{\Gamma_1}\,\hat{c}_{\text{in}}\\[2pt]
0\\[2pt]
\sqrt{\Gamma_2}\,\hat{b}_{\text{in}}
\end{bmatrix}.
\end{align}
Here, $\Gamma_{1}$ and $\Gamma_{2}$ are the external coupling rates at the in-plane and out-of-plane ports, respectively.  
The matrix $\hat{A}$ is diagonal, with entries
\(
A_{kk} = -\left( i\omega_k + \Gamma_k/2 \right)
\)
for $k=c,m,b$.  
The total decay rates of the in-plane and out-of-plane fields are $\Gamma_c = \Gamma_1 + \Gamma_{\text{i},c}$ and $\Gamma_b = \Gamma_2 + \Gamma_{\text{i},b}$, where $\Gamma_{\text{i},c}$ and $\Gamma_{\text{i},b}$ are internal cavity decay rates, and $\Gamma_m$ is the intrinsic magnon linewidth.

The input--output boundary conditions are written in terms of dimensionless quantities as
\begin{align}
\hat{a}_{\text{in}} + \hat{a}_{\text{out}} = \sqrt{\kappa_{j}}\,\hat{a},
\label{bound_cond_1}
\end{align}
where $\kappa_{j} G_c \equiv \Gamma_{j}$ (port $j = 1$ in-plane, $ j = 2$ out-of-plane) and $a=b,c$.  
With this convention, the average number of input and output photons in the time scale set by $G_c$ is given by
$\langle \hat{a}_{\text{in}}^\dagger \hat{a}_{\text{in}} \rangle = N_{\text{in}}$ and
$\langle \hat{a}_{\text{out}}^\dagger \hat{a}_{\text{out}} \rangle = N_{\text{out}}$.  
We do not fix $a=b$ or $a=c$ at this stage in order to keep the formalism general: depending on the specific experimental configuration, either the in-plane field, the out-of-plane field, or both may correspond to cavity modes (for instance, in Ref.~\cite{xu2020floquet} the in-plane field is a cavity mode whereas the out-of-plane field is not).

We now proceed by applying a one-sided Fourier transform to the equations of motion and taking expectation values.  
Since the input fields are prepared in coherent states, all expectation values can be evaluated using coherent states as well.  
We define the Fourier-transformed operators by
\(
\tilde{a}(\omega) = \mathcal{F}[\hat{a}(t)]
\)
and divide the equations by $G_c$ to obtain dimensionless coefficients.  
This leads to
\begin{align}
\begin{bmatrix}
\ell_c^- & -i        & 0 \\
-i       & \ell_m^-  & 0 \\
0        & 0         & \ell_b^-
\end{bmatrix}
\begin{bmatrix}
\langle \tilde{c} \rangle \\
\langle \tilde{m} \rangle \\
\langle \tilde{b} \rangle
\end{bmatrix}
=&\,
\frac{i}{2}\eta \epsilon^{-1}
\begin{bmatrix}
0\\[2pt]
0\\[2pt]
\delta
\end{bmatrix}
+
\begin{bmatrix}
\xi_c^-\\[2pt]
0\\[2pt]
\xi_b^-
\end{bmatrix}\nonumber\\[4pt]
&- i\eta \epsilon
\begin{bmatrix}
0\\[2pt]
B_m\\[2pt]
B_b
\end{bmatrix}
- i\epsilon^2
\begin{bmatrix}
C_c\\[2pt]
C_m\\[2pt]
0
\end{bmatrix},
\label{fourier_eqmot}
\end{align}
together with the Fourier-domain boundary conditions
\begin{align}
\langle \tilde{a}_{\text{in}} \rangle + \langle \tilde{a}_{\text{out}} \rangle
= \sqrt{\kappa_{j}}\,\langle \tilde{a} \rangle,
\end{align}
which follow the same conventions as the normalized decay  Eq.(\ref{bound_cond_1}). In addition, we have introduced $\eta = G_b/G_c$, and $\delta = \delta(\omega)$, and defined
\begin{align}
\ell_a^{\pm}(\omega)
&= \frac{\Gamma_a}{2G_c} - i\left( \frac{\omega \pm \omega_a}{G_c} \right)\nonumber\\
&\equiv \frac{\kappa_a}{2} - i\left( \frac{\omega \pm \omega_a}{G_c} \right),
\end{align}
for $a=\{c,m,b\}$, together with the spectral functions
\begin{align}
B_m(\omega) &\equiv \mathcal{F}\bigl[ \langle \hat{b}\hat{m} \rangle \bigr]
            + \mathcal{F}\bigl[ \langle \hat{b}^\dagger \hat{m} \rangle \bigr],\\
B_b(\omega) &\equiv \mathcal{F}\bigl[ \langle \hat{m}^\dagger \hat{m} \rangle \bigr],\\
C_c(\omega) &\equiv \frac{1}{2}\mathcal{F}\bigl[ \langle \hat{m}^\dagger \hat{m}^2 \rangle \bigr],\\
C_m(\omega) &\equiv 2\mathcal{F}\bigl[ \langle \hat{m}^\dagger \hat{m}\hat{c} \rangle \bigr]
              + \mathcal{F}\bigl[ \langle \hat{m}^2 \hat{c}^\dagger \rangle \bigr].
\end{align}

The coherent input fields yield Fourier transforms that are delta functions centered at the corresponding drive frequencies.  
Writing the input amplitudes as
\begin{align}
\xi_a^{\pm}(\omega)
\equiv \sqrt{\kappa_{j}}\,\zeta_a e^{i\theta_a} \delta(\omega \pm \omega_a)
\equiv \xi_a \delta(\omega \pm \omega_a),
\end{align}
we interpret $\zeta_a^2$ as the average number of photons in mode $a$ (with $a=c,b$) in the time scale set by $G_c$.

Finally, we propose the following perturbative expansion of the expectation values:
\begin{align}
\begin{bmatrix}
\langle \tilde{c} \rangle\\[2pt]
\langle \tilde{m} \rangle\\[2pt]
\langle \tilde{b} \rangle
\end{bmatrix}
=&\,
\epsilon^{-1}
\begin{bmatrix}
\langle \tilde{c}^{(-)} \rangle\\[2pt]
\langle \tilde{m}^{(-)} \rangle\\[2pt]
\langle \tilde{b}^{(-)} \rangle
\end{bmatrix}
+
\begin{bmatrix}
\langle \tilde{c}^{(0)} \rangle\\[2pt]
\langle \tilde{m}^{(0)} \rangle\\[2pt]
\langle \tilde{b}^{(0)} \rangle
\end{bmatrix}\nonumber\\[4pt]
&+
\epsilon
\begin{bmatrix}
\langle \tilde{c}^{(1)} \rangle\\[2pt]
\langle \tilde{m}^{(1)} \rangle\\[2pt]
\langle \tilde{b}^{(1)} \rangle
\end{bmatrix}
+
\epsilon^2
\begin{bmatrix}
\langle \tilde{c}^{(2)} \rangle\\[2pt]
\langle \tilde{m}^{(2)} \rangle\\[2pt]
\langle \tilde{b}^{(2)} \rangle
\end{bmatrix}
+ \dots,
\end{align}
and match powers of $\epsilon$ on both sides of Eq.~\eqref{fourier_eqmot} to obtain a hierarchy of equations for each perturbative order.  
The explicit derivations are summarized below, and more detailed steps are provided in Appendix~\ref{pert_calc}.

\subsection{Negative Order}

For the negative order to be consistent with the perturbative picture, we require
$\langle \tilde{m}^{(-)} \rangle = 0$.  
Physically, the magnon field is a response to the excitation field; in the absence of a linear drive, there should be no coherent magnon amplitude.  
This is analogous to perturbative schemes for the Lorentz oscillator in nonlinear optics, where the polarization at negative order vanishes in the absence of driving terms~\cite{boyd2008nonlinear}.  
With this assumption, it follows that $\langle \tilde{c}^{(-)} \rangle = 0$ and the only nontrivial contribution at negative order comes from $\langle \tilde{b}^{(-)} \rangle$, which satisfies
\begin{align}
\langle \tilde{b}^{(-)}(\omega) \rangle
= \frac{1}{2}\frac{i\eta\,\delta(\omega)}{\ell_b^-(\omega)}
= \frac{iG_b\,\delta(\omega)}{\Gamma_b + 2i\omega_b},
\label{neg_order}
\end{align}
where we used the identity $f(x)\delta(x-a) = f(a)\delta(x-a)$.  
Since the out-of-plane field couples to the magnon at order $\epsilon$, this negative-order contribution acts as a self-term that modifies the spectral equations for all higher orders of $\langle \tilde{m} \rangle$.

\subsection{Zeroth Order}

The presence of the negative-order solution induces a zeroth-order contribution to the magnon equation of motion.  
At zeroth order, we obtain
\begin{align}
\begin{bmatrix}
\ell_c^- & -i       & 0 \\
-i       & \ell_m^- & 0 \\
0        & 0        & \ell_b^-
\end{bmatrix}
\begin{bmatrix}
\langle \tilde{c}^{(0)} \rangle\\[2pt]
\langle \tilde{m}^{(0)} \rangle\\[2pt]
\langle \tilde{b}^{(0)} \rangle
\end{bmatrix}
=
\begin{bmatrix}
\xi_c^-\\[2pt]
-i\eta\,B_m^{(-)}\\[2pt]
\xi_b^-
\end{bmatrix},
\end{align}
where
\begin{align}
B_m^{(-)}(\omega)
&= \mathcal{F}\bigl[\langle \hat{b}^{(-)}\hat{m}^{(0)} \rangle \bigr]
 + \mathcal{F}\bigl[\langle \hat{b}^{(-)\dagger} \hat{m}^{(0)} \rangle \bigr]\nonumber\\
&= \left( \frac{iG_b}{\Gamma_b + 2i\omega_b}
      - \frac{iG_b}{\Gamma_b - 2i\omega_b} \right)
   \langle \tilde{m}^{(0)}(\omega) \rangle\nonumber\\
&= \frac{4G_b \omega_b}{\Gamma_b^2 + 4\omega_b^2}\,
   \langle \tilde{m}^{(0)}(\omega) \rangle.
\end{align}

In the underdamped regime, this correction slightly shifts the effective magnon frequency by
\(
\eta\,\frac{G_b}{\omega_b} = \frac{G_b^2}{G_c \omega_b}
\equiv \Delta_b/G_c.
\) We note that the quantization of the modulation mode is not directly responsible for the appearance of the magnon frequency shift. What is required is the inclusion of the ground-state Zeeman term because it results in an additional, effective static magnetic field within the ferromagnet. We justify this argument in Appendix \ref{app_D}.

Recalling that we normalized all coefficients by $G_c$, the magnon factor becomes
\begin{align}
\ell_m^-
= \frac{\kappa_m}{2} - i\left( \frac{\omega - \omega_m}{G_c} \right)
\rightarrow
\frac{\kappa_m}{2} - i\left( \frac{\omega - \omega_m}{G_c} \right)
+ i\frac{\Delta_b}{G_c}\nonumber\\
= \frac{\kappa_m}{2}
- i\left[ \frac{\omega - (\omega_m + \Delta_b)}{G_c} \right]
\equiv \ell_{m,\Delta}^-.
\end{align}
Thus, the negative-order term can be absorbed into a renormalized magnon frequency $\omega_m \rightarrow \omega_m + \Delta_b$.

Solving the zeroth-order system with this redefinition yields
\begin{align}
\begin{bmatrix}
\langle \tilde{c}^{(0)} \rangle\\[2pt]
\langle \tilde{m}^{(0)} \rangle
\end{bmatrix}
&=
\frac{1}{1 + \ell_{m,\Delta}^- \ell_c^-}
\begin{bmatrix}
\ell_{m,\Delta}^- & i\\[2pt]
i                & \ell_c^-
\end{bmatrix}
\begin{bmatrix}
\xi_c^-\\[2pt]
0
\end{bmatrix}\nonumber\\[4pt]
&\equiv
D^-(\omega)
\begin{bmatrix}
\ell_{m,\Delta}^-\,\xi_c^-\\[2pt]
i\,\xi_c^-
\end{bmatrix},
\end{align}
with
\begin{align}
\langle \tilde{b}^{(0)}(\omega) \rangle
= F_b^-(\omega)\,\xi_b^-,
\qquad
F_b^-(\omega) \equiv \frac{1}{\ell_b^-(\omega)},
\label{Fb_def}
\end{align}
and
\begin{align}
    D^-(\omega) \equiv \dfrac{1}{1+\ell_{m,\Delta}^-\ell_c^-}.
    \label{D_def}
\end{align}

In practice, we drop the subscript $\Delta$ and understand $\ell_m^-$ to include the shift $\Delta_b$.

\subsection{First Order}

The first-order spectra do not produce visible features in the reflectance by themselves.  
However, they enter the second-order equations and are responsible for generating the Floquet sidebands observed experimentally.  
At first order we obtain
\begin{align}
\begin{bmatrix}
\ell_c^- & -i       & 0 \\
-i       & \ell_m^- & 0 \\
0        & 0        & \ell_b^-
\end{bmatrix}
\begin{bmatrix}
\langle \tilde{c}^{(1)} \rangle\\[2pt]
\langle \tilde{m}^{(1)} \rangle\\[2pt]
\langle \tilde{b}^{(1)} \rangle
\end{bmatrix}
=
-i\eta
\begin{bmatrix}
0\\[2pt]
B_m^{(0)}\\[2pt]
B_b^{(0)}
\end{bmatrix},
\end{align}
where
\begin{align}
B_m^{(0)}(\omega)
&\equiv \mathcal{F}\bigl[ \langle \hat{b}^{(0)}\hat{m}^{(0)} \rangle \bigr]
 + \mathcal{F}\bigl[ \langle \hat{b}^{(0)\dagger}\hat{m}^{(0)} \rangle \bigr],\\
B_b^{(0)}(\omega)
&\equiv \mathcal{F}\bigl[ \langle \hat{m}^{(0)\dagger}\hat{m}^{(0)} \rangle \bigr].
\end{align}
Solving the $c$--$m$ subsystem yields
\begin{align}
\begin{bmatrix}
\langle \tilde{c}^{(1)} \rangle\\[2pt]
\langle \tilde{m}^{(1)} \rangle
\end{bmatrix}
&=
-\frac{i\eta}{1+\ell_m^- \ell_c^-}
\begin{bmatrix}
\ell_m^- & i\\[2pt]
i        & \ell_c^-
\end{bmatrix}
\begin{bmatrix}
0\\[2pt]
B_m^{(0)}
\end{bmatrix}\nonumber\\[4pt]
&=
-i\eta\,D^-(\omega)\,B_m^{(0)}(\omega)
\begin{bmatrix}
i\\[2pt]
i\ell_c^-
\end{bmatrix},
\end{align}
and for the out-of-plane mode
\begin{align}
\langle \tilde{b}^{(1)}(\omega) \rangle
= -i\eta\,B_b^{(0)}(\omega)\,F_b^-(\omega).
\end{align}
The explicit forms of $B_m^{(0)}$ and $B_b^{(0)}$ are worked out in Appendix~\ref{pert_calc}, and can be written as
\begin{align}
B_m^{(0)}(\omega)
&= \xi_b F_b(\omega_b)\,\langle \tilde{m}^{(0)}(\omega-\omega_b) \rangle \nonumber\\
&\quad + \xi_b^* F_b^*(\omega_b)\,\langle \tilde{m}^{(0)}(\omega+\omega_b) \rangle,
\end{align}
and
\begin{align}
B_b^{(0)}(\omega)
= -i\,\xi_c^* D^{-*}(\omega_c)\,\langle \tilde{m}^{(0)}(\omega+\omega_c) \rangle,
\end{align}
where $F_b^-(\omega_b)$ and $D^-(\omega)$ are defined in Eq.(\ref{Fb_def}) and Eq.($\ref{D_def}$), respectively.

\subsection{Second Order}

The second-order spectra contain both the Floquet sidebands and the contributions from the second-order terms in the Holstein--Primakoff transformation.  
In most experiments, the in-plane cavity drive is weak enough that the coherently-driven magnon population remains close to unity, and HP corrections are therefore small.  
Nonetheless, for completeness we retain them in our analysis.

At second order we obtain
\begin{align}
\begin{bmatrix}
\ell_c^- & -i       & 0 \\
-i       & \ell_m^- & 0 \\
0        & 0        & \ell_b^-
\end{bmatrix}
\begin{bmatrix}
\langle \tilde{c}^{(2)} \rangle\\[2pt]
\langle \tilde{m}^{(2)} \rangle\\[2pt]
\langle \tilde{b}^{(2)} \rangle
\end{bmatrix}
=&\,
-i\eta
\begin{bmatrix}
0\\[2pt]
B_m^{(1)}\\[2pt]
B_b^{(1)}
\end{bmatrix}
-\frac{i}{2}
\begin{bmatrix}
C_c^{(0)}\\[2pt]
C_m^{(0)}\\[2pt]
0
\end{bmatrix},
\end{align}
with
\begin{align}
B_m^{(1)}(\omega)
&\equiv \mathcal{F}\bigl[ \langle \hat{b}^{(1)}\hat{m}^{(0)} \rangle \bigr]
   + \mathcal{F}\bigl[ \langle \hat{b}^{(1)\dagger}\hat{m}^{(0)} \rangle \bigr]\nonumber\\
&\quad + \mathcal{F}\bigl[ \langle \hat{b}^{(0)}\hat{m}^{(1)} \rangle \bigr]
   + \mathcal{F}\bigl[ \langle \hat{b}^{(0)\dagger}\hat{m}^{(1)} \rangle \bigr],\\
B_b^{(1)}(\omega)
&\equiv \mathcal{F}\bigl[ \langle \hat{m}^{(1)\dagger}\hat{m}^{(0)} \rangle \bigr]
   + \mathcal{F}\bigl[ \langle \hat{m}^{(0)\dagger}\hat{m}^{(1)} \rangle \bigr],
\end{align}
and
\begin{align}
C_c^{(0)}(\omega)
&\equiv \mathcal{F}\bigl[ \langle \hat{m}^{(0)\dagger}\hat{m}^{(0)2} \rangle \bigr],\\
C_m^{(0)}(\omega)
&\equiv 2\mathcal{F}\bigl[ \langle \hat{m}^{(0)\dagger}\hat{m}^{(0)}\hat{c}^{(0)} \rangle \bigr]
  + \mathcal{F}\bigl[ \langle \hat{m}^{(0)2}\hat{c}^{(0)\dagger} \rangle \bigr].
\end{align}
We write
\begin{align}
B_m^{(1)}(\omega)
= B_m^{(1),m}(\omega) + B_m^{(1),b}(\omega),
\end{align}
where
\begin{align}
B_m^{(1),m}(\omega)
=& -\frac{i}{2}\eta |\xi_b|^2 |F_b(\omega_b)|^2 D^-(\omega-\omega_b)\nonumber\\
&\times \ell_c(\omega-\omega_b)\,\langle \tilde{m}^{(0)}(\omega) \rangle \nonumber\\
&-\frac{i}{2}\eta |\xi_b|^2 |F_b(\omega_b)|^2 D^-(\omega+\omega_b)\nonumber\\
&\times \ell_c(\omega+\omega_b)\,\langle \tilde{m}^{(0)}(\omega) \rangle,
\end{align}
and
\begin{align}
B_m^{(1),b}(\omega)
= -i|\xi_c|^2 \bigl[ F_b(0) - F_b^*(0) \bigr]
  \langle \tilde{m}^{(0)}(\omega) \rangle.
\end{align}
The remaining functions read
\begin{align}
C_c^{(0)}(\omega)
&= |\xi_c|^2 |D^-(\omega_c)|^2 \langle \tilde{m}^{(0)}(\omega) \rangle,\\
C_m^{(0)}(\omega)
&= -i|\xi_c|^2 |D^-(\omega_c)|^2
   \bigl[ 2\ell_m(\omega_c) - \ell_m^*(\omega_c) \bigr]
   \langle \tilde{m}^{(0)}(\omega) \rangle,
\end{align}
with $\langle \tilde{m}^{(0)}(\omega) \rangle = iD^-(\omega)\langle \tilde{c}_{\text{in}}(\omega) \rangle$.  
Solving the second-order equations for the in-plane field, we obtain
\begin{align}
\langle \tilde{c}^{(2)}(\omega) \rangle
=&\, \frac{1}{2}D^-(\omega)
\bigl[ \eta B_m^{(1)}(\omega) - i\ell_m(\omega) C_c^{(0)}(\omega) \bigr]\nonumber\\
&+ \frac{1}{2}D^-(\omega)\,C_m^{(0)}(\omega),
\end{align}
which provides the second-order contribution to the reflectance and encodes both the Floquet sidebands and HP corrections.

\section{Reflectance Spectra\label{discussion}}

In two-port microwave cavity setups, the reflection coefficient $S_{11}(\omega)$ can be obtained by sending a sinusoidal input signal of frequency $\omega$ into the cavity through a certain port (let us call it port $1$, corresponding to the $\hat{c}$ field input) and observing the amplitude and/or phase of the output signal at the same port. A common device that is used to measure both amplitude and phase in microwave cavity setups is the Vector Network Analyzer (VNA), which allows for a direct measurement of the absolute reflectance or transmittance.

For completeness, the transmittance coefficient $S_{21}(\omega)$ can be obtained by sending an input signal of frequency $\omega$ into the cavity through port $1$ and observing the output signal at port $2$ (which would correspond to the $\hat{b}$ field input). 
\begin{align}
    S_{11}(\omega)=\dfrac{\braket{\tilde{c}_{\text{out}}(\omega)}}{\braket{\tilde{c}_{\text{in}}(\omega)}}=-1+\sqrt{\kappa_1}\dfrac{\braket{\tilde{c}(\omega)}}{\braket{\tilde{c}_{\text{in}}(\omega)}},
\end{align}\begin{align}
     S_{21}(\omega)=\dfrac{\braket{\tilde{b}_{\text{out}}(\omega)}}{\braket{\tilde{c}_{\text{in}}(\omega)}}=-\dfrac{\braket{\tilde{b}_{\text{in}}(\omega)}}{\braket{\tilde{c}_{\text{in}}(\omega)}}+\sqrt{\kappa_2}\dfrac{\braket{\tilde{b}(\omega)}}{\braket{\tilde{c}_{\text{in}}(\omega)}}.
\end{align}

Given that the $\hat{c}$ and $\hat{b}$ fields are very detuned from one another ($\omega_c=8.5 \text{ GHz}$, $\omega_b\sim 10 \text{ MHz}$), we effectively treat the physical situation as if we had two one-port cavities coupled by the uniform magnon mode. 

\subsection{Second-Order Reflectance\label{sec_ord}}

In this section we evaluate the second-order reflectance of the in-plane field using the experimental parameters reported in Ref.~\cite{xu2020floquet}.  
In typical cavity magnonics experiments, the reflectance (or probe) field is the in-plane mode $\hat{c}$, and the reflection coefficient is defined as
\begin{align}
S_{11}(\omega)
&= \frac{\langle \tilde{c}_{\text{out}}(\omega)\rangle}
        {\langle \tilde{c}_{\text{in}}(\omega)\rangle}
= -1 +\sqrt{\kappa_1} \frac{\langle \tilde{c}(\omega)\rangle}
             {\langle \tilde{c}_{\text{in}}(\omega)\rangle} \nonumber\\
&= S_{11}^{(0)}(\omega) + \epsilon^2 S_{11}^{(2)}(\omega) + \dots,
\label{refl_def}
\end{align}
where
\begin{align}
S_{11}^{(0)}(\omega)
&= -1 + \sqrt{\kappa_1}\frac{\langle \tilde{c}^{(0)}(\omega)\rangle}
                {\langle \tilde{c}_{\text{in}}(\omega)\rangle},\\[3pt]
S_{11}^{(2)}(\omega)
&=\sqrt{\kappa_1} \frac{\langle \tilde{c}^{(2)}(\omega)\rangle}
        {\langle \tilde{c}_{\text{in}}(\omega)\rangle}.
\end{align}
The zeroth-order term describes the standard cavity--magnon hybridization, while the second-order correction encodes the effect of the Floquet modulation and the higher-order magnon terms on the frequencies absorbed and reflected by the cavity.

\subsection{Experimental Parameters and Validity of the Expansion}

For comparison with experiment, we adopt the parameters of Ref.~\cite{xu2020floquet}.  
The ferromagnet is a YIG sphere \cite{serga2010yig} of radius $R = 200~\mu\mathrm{m}$.  
Using the lattice parameter $a = 1.25~\mathrm{nm}$~\cite{cherepanov1993saga} and spin $s=5/2$ per atomic site, we estimate the perturbative parameter as
\[
\epsilon = (2Ns)^{-1/2} \approx 3.414\times 10^{-9}.
\]
The natural magnon linewidth is taken as $\Gamma_m/2\pi = 4.4~\mathrm{MHz}$.

For the in-plane (cavity) field we use
$\omega_c/2\pi = 8.5~\mathrm{GHz}$,
spin-enhanced coupling
$G_c/2\pi = 14.0~\mathrm{MHz}$, and total decay rate
$\Gamma_c/2\pi = 2.0~\mathrm{MHz}$.  
For the out-of-plane (modulation) field we take
$\omega_b/2\pi$ in the range $0$--$40.0~\mathrm{MHz}$ and
$\Gamma_b/2\pi = \Gamma_2/2\pi = 5.6~\mathrm{MHz}$.

To make the connection with the perturbative expansion explicit, we introduce the dimensionless parameters
\begin{align}
u_c &= \frac{\epsilon\,\xi_c}{\kappa_1},\qquad
u_b = \frac{\epsilon\,\eta\,\xi_b}{\kappa_b},
\end{align}
which measure the normalized photon fluxes in the in-plane and out-of-plane modes, respectively.  
In Sec.~\ref{pert_pam} we relate these parameters to the semi-classical Floquet coupling $\Omega$ and show how they can be expressed in terms of more fundamental quantities.

As in any perturbative scheme, it is crucial to identify the regime of validity of the expansion and thereby constrain $u_b$ and $u_c$.  
We do this by demanding that the reflectance remain physical, i.e., that the magnitude of the reflection coefficient at the hybridized cavity frequencies does not exceed unity:
\begin{align}
|S_{11}(\omega_c \pm G_c)| \leq 1.
\end{align}
Violation of this condition signals that higher-order contributions outside our truncation become important and the second-order approximation ceases to be reliable.

\begin{figure}[ht]
    \centering
    \includegraphics[width=1\linewidth]{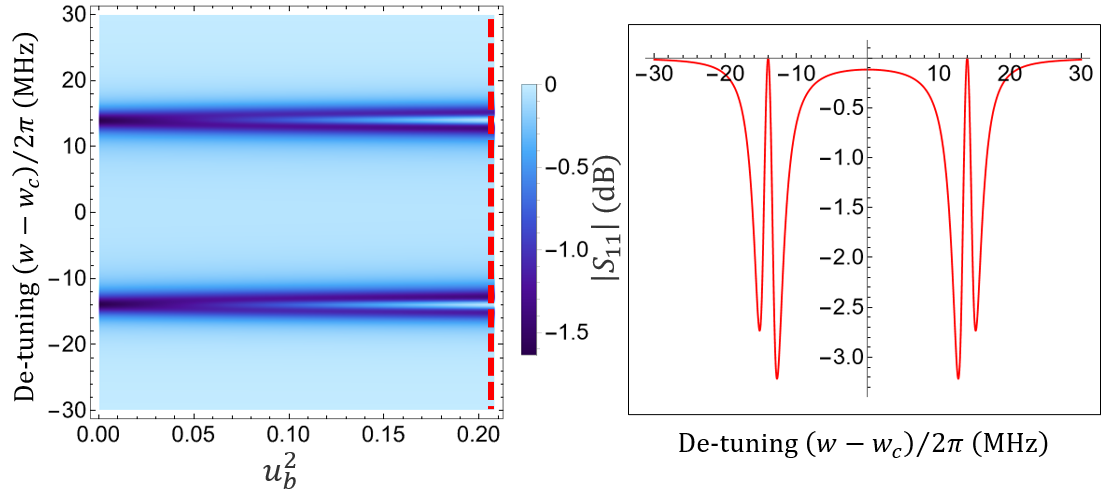}
    \caption{(Left) Absolute reflectance $|S_{11}|$ in dB as a function of the in-plane detuning $(\omega-\omega_c)/2\pi$ and the squared modulation parameter $u_b^2$. (Right) Reflectance at the threshold of validity, $u_b = u_b^{\mathrm{max}}$.}
    \label{fig:w_x_xib}
\end{figure}

We first consider the case $u_c=0$ and vary $u_b$.  
Figure~\ref{fig:w_x_xib} (left) shows the absolute reflectance $|S_{11}|$ in dB as a function of the in-plane detuning $(\omega-\omega_c)/2\pi$ and the squared modulation parameter $u_b^2$.  
The threshold value $u_b^{\text{max}}$ is defined as the value of $u_b$ at which the reflectance at the hybrid mode frequencies, $\omega_c \pm G_c$, first exceeds unity.  
Using the parameters above, we find
\[
u_b^{\text{max}} = 0.456.
\]
The right panel of Fig.~\ref{fig:w_x_xib} shows a cut at this threshold.

\begin{figure}[ht]
    \centering
    \includegraphics[width=1\linewidth]{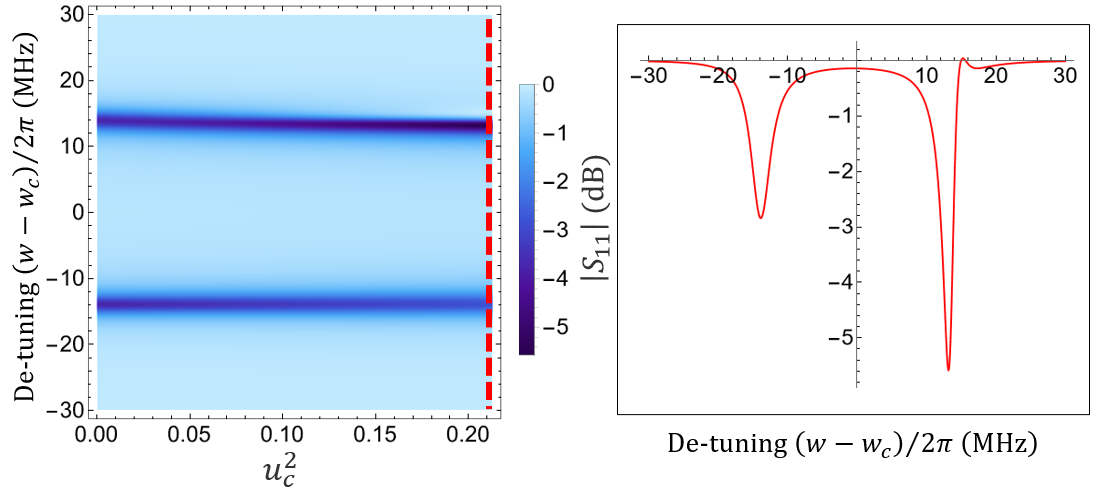}
    \caption{(Left) Absolute reflectance $|S_{11}|$ in dB as a function of the in-plane detuning $(\omega-\omega_c)/2\pi$ and the squared in-plane parameter $u_c^2$. (Right) Reflectance at the threshold of validity, $u_c = u_c^{\mathrm{max}}$.}
    \label{fig:w_x_xic}
\end{figure}

Next, we set $u_b=0$ and analyze the dependence on $u_c$.  
Increasing $u_c$ enhances the absorption near the upper hybrid mode, as shown in Fig.~\ref{fig:w_x_xic}.  
Here, we define the threshold $u_c^{\text{max}}$ as the value of $u_c$ for which the reflectance at the shifted upper hybrid frequency, $\omega_c + G_c + \Delta_c$, reaches unity, with $\Delta_c/2\pi = 1.0~\mathrm{MHz}$ in our numerics.  
This yields
\[
u_c^{\text{max}} = 0.465.
\]

In typical cavity magnonics setups the in-plane drive is kept weak because of its strong coupling to the magnon mode.  
Accordingly, in the following we set $u_c = 10^{-6} \approx 0$, so that distortions due to a large cavity drive are negligible and the modulation-induced features are isolated.

For reference, the main parameters used in our calculations are summarized in Table~\ref{tab:params}.

\begin{table}[t]
\caption{Representative parameters used in the reflectance calculations, largely based on Ref.~\cite{xu2020floquet}.}
\label{tab:params}
\begin{ruledtabular}
\begin{tabular}{lcc}
Quantity & Symbol & Value \\
\hline
YIG sphere radius & $R$ & $200~\mu\mathrm{m}$\\
Lattice parameter & $a$ & $1.25~\mathrm{nm}$\\
Spin per site & $s$ & $5/2$\\
Perturbative parameter & $\epsilon$ & $3.414\times 10^{-9}$\\
Magnon linewidth & $\Gamma_m/2\pi$ & $4.4~\mathrm{MHz}$\\
Cavity frequency & $\omega_c/2\pi$ & $8.5~\mathrm{GHz}$\\
Cavity--magnon coupling & $G_c/2\pi$ & $14.0~\mathrm{MHz}$\\
Cavity decay rate & $\Gamma_c/2\pi$ & $2.0~\mathrm{MHz}$\\
Modulation frequency & $\omega_b/2\pi$ & $0$--$40.0~\mathrm{MHz}$\\
Modulation decay rate & $\Gamma_b/2\pi$ & $5.6~\mathrm{MHz}$\\
 USFC modulation decay rate& &$0.365~\mathrm{MHz}$\\
Threshold (modulation) & $u_b^{\mathrm{max}}$ & $0.456$\\
Threshold (cavity) & $u_c^{\mathrm{max}}$ & $0.465$\\
\end{tabular}
\end{ruledtabular}
\end{table}

\subsection{Reflectance Plots}

\begin{figure}[ht]
    \centering
    \includegraphics[width=1\linewidth]{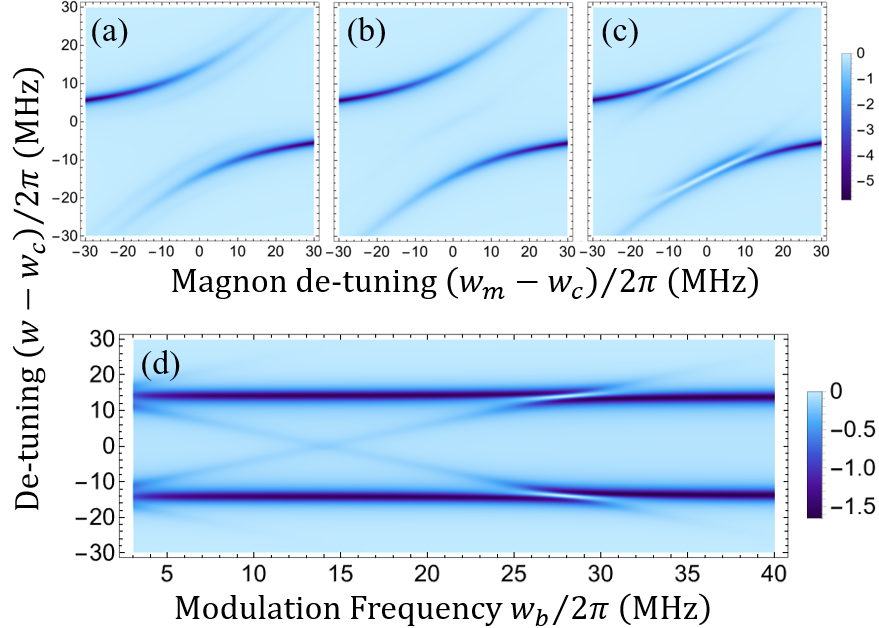}
    \caption{Second-order reflectance spectra. (a)--(c) Detuning of the cavity mode vs.\ magnon detuning for modulation frequencies $\omega_b/2\pi =$ (a) $5~\mathrm{MHz}$, (b) $14~\mathrm{MHz}$, and (c) $28~\mathrm{MHz}$. (d) Reflectance as a function of cavity detuning and modulation frequency.}
    \label{refl_graphs}
\end{figure}

To maximize the visibility of the Floquet sidebands within the perturbative regime, we set $u_b = u_b^{\text{max}} = 0.456$ and $u_c \approx 0$.  
The resulting reflectance spectra are shown in Fig.~\ref{refl_graphs}.  
Panels (a)--(c) display $|S_{11}|$ as a function of the cavity detuning and magnon detuning for modulation frequencies $\omega_b/2\pi = 5~\mathrm{MHz}$, $14~\mathrm{MHz}$, and $28~\mathrm{MHz}$, respectively.  
Panel (d) shows the reflectance as a function of the cavity detuning and the modulation frequency.  
The overall pattern of sidebands and hybrid-mode features closely resembles the dominant spectral characteristics observed in Ref.~\cite{xu2020floquet}, demonstrating that the second-order perturbative theory captures the essential physics of the Floquet cavity magnonics experiment.

We now turn to the ultra-strong Floquet coupling (USFC) regime.  This regime occurs when the level splitting between the hybrid magnon polariton modes becomes comparable or lesser than the Autler-Townes Splitting ($2G_c \lesssim \Delta \omega _{\text{AT}}$). Using general Floquet Theory, this regime marks the failure of the RWA for the out-of-plane field and requires the inclusion of counterrotating terms of hybrid magnon-polariton field operators \cite{xu2020floquet,PhysRev.138.B979}. In our perturbative method, these counterrotating contributions are automatically included but expressed in terms of the un-hybridized field operators of the cavity and the uniform magnon modes ($\hat{c},\hat{m}$).

\begin{figure}[ht]
    \centering
    \includegraphics[width=1\linewidth]{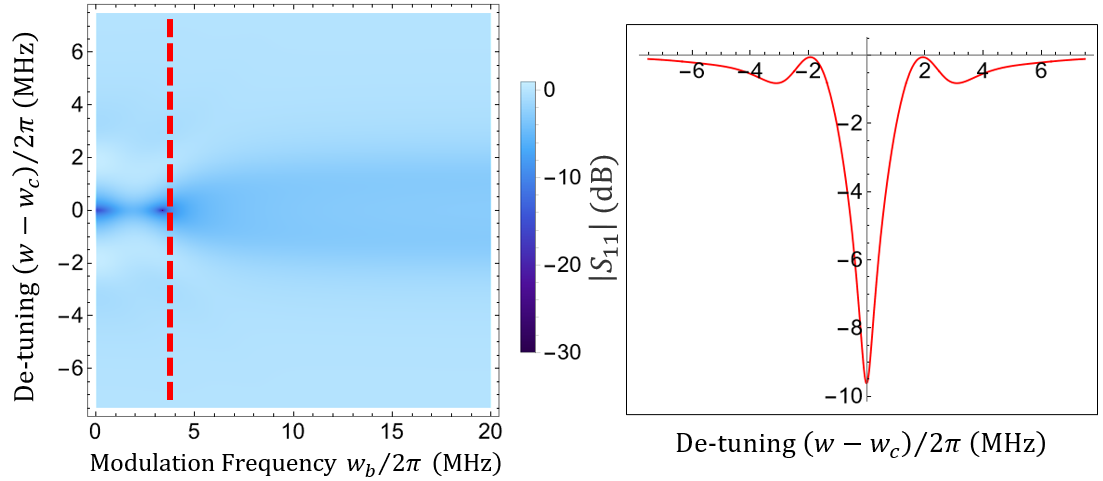}
    \caption{Ultra-strong Floquet coupling regime. (Left) Reflectance as a function of cavity detuning and modulation frequency $\omega_b/2\pi$. (Right) Reflectance as a function of detuning for fixed $\omega_b/2\pi = 3.85~\mathrm{MHz}$, showing a pronounced reflection dip.}
    \label{w_x_wb(USFC)}
\end{figure}

For the simulations respective to this regime, we set $G_c/2\pi = 1.825~\mathrm{MHz}$ and $\Gamma_c/2\pi = 2.6~\mathrm{MHz}$.  
For the modulation field we consider $\omega_b/2\pi$ in the range $0$--$20.0~\mathrm{MHz}$ and $\Gamma_b/2\pi = 0.365~\mathrm{MHz}$.  
With these values, the perturbative threshold is found to be
$u_b^{\text{max}} = 3.79$, and we again use this value to maximize the visibility of the modulation-induced features.  
The resulting spectra are shown in Fig.~\ref{w_x_wb(USFC)}.  
The spectral profile is qualitatively different from the previous case and exhibits pronounced ``dips'' in the reflectance at certain frequencies.

\begin{figure}[ht]
    \centering
    \includegraphics[width=1\linewidth]{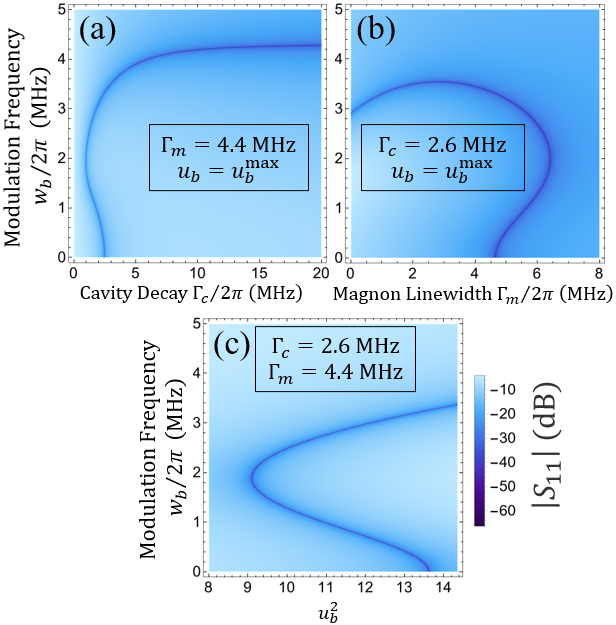}
    \caption{Absolute reflection dip at the central frequency $\omega = \omega_c = \omega_m$ as a function of (a) the cavity decay rate $\Gamma_c/2\pi$, (b) the magnon linewidth $\Gamma_m/2\pi$, and (c) the squared modulation parameter $u_b^2$ in the ultra-strong Floquet coupling regime.}
    \label{USFC_graphs}
\end{figure}

The location and depth of these reflectance dips are sensitive to the system parameters, as illustrated in Fig.~\ref{USFC_graphs}, which shows the dip at the central frequency $\omega = \omega_c = \omega_m$ as a function of (a) the total cavity decay rate $\Gamma_c/2\pi$, (b) the magnon linewidth $\Gamma_m/2\pi$, and (c) the squared modulation parameter $u_b^2$.  
This sensitivity suggests that such dips could serve as a diagnostic feature for extracting or constraining system parameters in strongly driven cavity magnonics experiments.

\subsection{Modulation Field Volume Factor $\alpha$}

There is an important conceptual difference between the derivations of the in-plane and out-of-plane couplings.  For the in-plane field, the relevant frequency is the cavity frequency $\omega_c$ and the electromagnetic energy is approximately distributed throughout the volume $V_c$ occupied by the cavity mode. In Ref.\cite{tabuchi2016quantum}, $V_c=22\times 18\times 3\text{ mm}^3=1188\text{ mm}^3$ is the entire cavity volume. In Ref.\cite{xu2020floquet}, the field lines are concentrated within a cylindrical Dielectric Resonator with diameter $6\text{ mm}$ and height of $4\text{ mm}$, such that $V_c\approx 113.1\text{ mm}^3$. Within our model, this leads to the estimate
\begin{align}
\frac{g_c}{2\pi}=\dfrac{\gamma}{2}\sqrt{\dfrac{\mu_0\hbar \omega_c}{2V_c}}
\approx 78.4~\mathrm{mHz},
\end{align}
where the factor of $1/2$ arises from the different definitions of the spin-enhanced couplings for the in-plane and out-of-plane fields. This value has similar order of magnitude of the intrinsic in-plane coupling presented in Ref.\cite{tabuchi2016quantum} ($\sqrt{2s}g_c/2\pi \sim 40\text{ mHz}$).

In contrast, the modulation field is not confined by the cavity, and its effective volume is not uniquely defined.  
To model this more realistically, we introduce a phenomenological scaling factor $\alpha$ and write the effective modulation volume as
\begin{align}
V_b \equiv \alpha V_{\mathrm{f}},
\end{align}
corresponding to the region where the out-of-plane field is strongest in the experimental geometry of Ref.~\cite{xu2020floquet}.  
Within our framework, knowledge of the ferromagnet dimensions and a measurement of the modulation-induced magnon detuning $\Delta_b$ can, in principle, be used to estimate $\alpha$.

\subsection{Energy Shift Measurement}

From the negative-order contribution in the perturbative expansion, we found that the magnon frequency is shifted by\begin{align}
\Delta_b
= \frac{4G_b^2 \omega_b}{\Gamma_b^2 + 4\omega_b^2}.
\end{align}

Using the microscopic expression for the spin-enhanced coupling of Eq.(\ref{g_def}) and $V_b=\alpha Na^3$, we find
\begin{align}
G_b^2
= \frac{\gamma^2 \mu_0 s \hbar}{\alpha a^3} \omega_b\approx \left(2\pi \times  \dfrac{837.3}{\alpha }~\mathrm{MHz}\right)\omega_b,
   \label{Gb_wb}
\end{align}
and obtain
\begin{align}
   \Delta_b=\dfrac{\tilde{\Delta}_b}{\left(\dfrac{\Gamma_b}{2\omega_b}\right)^2+1},
\end{align}
where we define $\tilde{\Delta}_b=G_b^2/\omega_b=(2\pi \times 837.3\text{ MHz})\alpha^{-1}$. $\tilde{\Delta}_b$ is precisely the Zeeman interaction between the out-of-plane field and the internal magnetization field divided by the volume scaling factor $\alpha$
\begin{align}
\tilde{\Delta}_b
= \gamma \mu_0 \frac{\gamma s\hbar}{\alpha a^3}
= \gamma \mu_0 \frac{\gamma Ns\hbar}{\alpha V_{\mathrm{f}}}
=\dfrac{ \gamma \mu_0 M_s}{\alpha},
\label{delta_b_Magnetization}
\end{align}
where $V_{\mathrm{f}} = N a^3$ is the volume of the ferromagnet and $M_s$ is the modulus of the saturation magnetization. Using the modulation frequency interval of Fig.\ref{refl_graphs}(d) and $\Gamma_b/2\pi =5.6\text{ MHz}$, we find that the product between the energy shift and the phenomenological volume factor $\alpha$ should be restricted to the range
\begin{align}
  447.5~\text{MHz} \leq \dfrac{\alpha \Delta_b}{2\pi }\leq  833.2~\text{MHz}.
\end{align}
An interesting limit to note is in the underdamped regime ($\omega_b\gg \Gamma_b/2 $) whereby $\Delta_b\rightarrow \tilde{\Delta}_b$. 

In addition to the energy shift, the relation in Eq.(\ref{Gb_wb}) suggests the scaling law
\begin{align}
G_b = \sqrt{\tilde{\Delta}_b\omega_b}=\sqrt{\dfrac{2\pi \omega_b}{\alpha}}\,\times 28.94~\mathrm{MHz}^{1/2},
\label{spin_coupling_scaling}
\end{align}
which can be used as a convenient rule-of-thumb for estimating the spin-enhanced coupling from the modulation frequency and the effective volume scaling factor $\alpha$. In the experimental data of Ref.~\cite{xu2020floquet}, the magnon frequency is not measured directly, but inferred from the position of an external magnet generating the static field.  
For $\omega_b/2\pi \approx 10~\mathrm{MHz}$, our scaling yields $G_b/2\pi \approx 91.5~\mathrm{MHz}$.  
Using $\epsilon = 3.414\times 10^{-9}$ and $\alpha \approx 1$, this corresponds to an intrinsic out-of-plane magnetic-dipole coupling
$g_b/2\pi \approx 312.4~\mathrm{mHz}$.

\subsection{Semi-Classical Floquet Coupling $\Omega$\label{pert_pam}}

In the reflectance plots presented above, we have used $\Gamma_b \sim 0.4G_c$ for $G_c/2\pi = 14~\mathrm{MHz}$ and $\Gamma_b \sim 0.2G_c$ for $G_c/2\pi = 1.825~\mathrm{MHz}$.  
These choices were heuristic and, in particular, the threshold $u_b^{\text{max}}$ is independent of $\Gamma_b$.  
Since the modulation mode is not a cavity mode in the usual sense, its decay rate does not share the same physical origin as $\Gamma_c$.  
We can, however, obtain a more systematic estimate by connecting the quantum description to the semi-classical Floquet picture.

In the semi-classical approach, the modulation field is described as a classical coherent drive.  
The Floquet interaction is commonly written as
\begin{align}
\hat{V}_F
= 2\epsilon G_b \zeta_b \hat{m}^\dagger\hat{m}\cos(\omega_b t)
\;\leftrightarrow\;
\Omega\,\hat{m}^\dagger\hat{m}\cos(\omega_b t),
\end{align}
where $\zeta_b = \sqrt{N_b} = \xi_b/\sqrt{\kappa_b}$ is related to the mean photon number $N_b$, and $\Omega$ is the semi-classical coupling parameter between the magnon and the modulation field.  
Throughout, we have assumed $\kappa_b = \kappa_2$, i.e., that the total decay rate of the modulation mode equals the readout rate (no internal losses).  
Using $u_b = \epsilon \eta \xi_b/\kappa_b$ we find
\begin{align}
\Omega = 2G_c \sqrt{\kappa_b}\,u_b,
\end{align}
which can be inverted to give
\begin{align}
\Gamma_b
= \frac{1}{G_c}\left( \frac{\Omega}{2u_b} \right)^2.
\label{omega_gam}
\end{align}
This expression provides a direct connection between the semi-classical Floquet coupling $\Omega$ and our effective decay rate $\Gamma_b$.

For $G_c/2\pi = 14~\mathrm{MHz}$ and $u_b \leq u_b^{\text{max}} = 0.456$, using $\Omega/2\pi = 8~\mathrm{MHz}$ from Ref.~\cite{xu2020floquet} yields the bound
\[
\Gamma_b/2\pi \gtrsim 5.52~\mathrm{MHz}.
\]
Similarly, for $G_c/2\pi = 1.825~\mathrm{MHz}$ we obtain
\[
\Gamma_b/2\pi \gtrsim 0.044~\mathrm{MHz}.
\]
A more precise determination of $\Gamma_b$ would, of course, require detailed experimental characterization.  
Here, our goal is to demonstrate that the perturbative input--output framework naturally accommodates the semi-classical Floquet coupling and provides a scalable alternative to fully numerical treatments, particularly in more complex situations involving multiple quantum systems coupled to one or more ferromagnets inside a cavity.

\section{Conclusions}\label{conc}

We have developed a perturbative input--output formalism for cavity magnonics systems subject to a Floquet modulation and applied it to compute the reflectance spectra measured in recent experiments.  
By exploiting the hierarchy between intrinsic and spin-enhanced magnetic-dipole couplings, we identified the small parameter $\epsilon=(2Ns)^{-1/2}$ and systematically organized the cavity and magnon fields in powers of $\epsilon$.  
This allowed us to derive compact analytic expressions for the spectral response up to second order, providing a physically transparent alternative to fully numerical treatments of driven cavity magnonics.

Within this framework, we showed that the second-order correction to the in-plane reflectance reproduces the main features of the Floquet sideband structure observed in Ref.~\cite{xu2020floquet}, both in the standard and in the ultra-strong Floquet coupling regimes.  
In the latter case, our theory predicts pronounced reflection dips whose position and depth are highly sensitive to the cavity decay rate, magnon linewidth, and modulation strength, suggesting that such features can be used as diagnostics to extract or constrain system parameters in strongly driven cavity magnonics experiments.

A key conceptual result of our analysis is the identification of a modulation-induced magnon energy shift arising from the Zeeman interaction between the modulation field and the fully polarized ground state.  
This term, usually neglected in cavity magnonics models, appears at negative order in the perturbative expansion and leads to a sizable detuning of order $0.8~\mathrm{GHz}$ for realistic parameters.  
We showed that this shift can be expressed in terms of the saturation magnetization of the ferromagnet and interpreted as a measurable, previously overlooked contribution that must be taken into account in precision spectroscopic studies and in the design of Floquet-based magnonic devices.

We also introduced a phenomenological volume factor for the modulation field and connected the quantum description to the semi-classical Floquet coupling $\Omega$, thereby establishing a direct relation between $\Omega$ and the effective decay rate $\Gamma_b$ of the modulation mode.  
This connection clarifies how experimental control over the modulation amplitude and frequency maps onto the perturbative parameters $u_b$ and $u_c$, and delineates the regime of validity of the expansion in practical settings.

Because the perturbative expansion explicitly tracks the role of the spin-enhanced couplings, the method is straightforward to generalize to more complex hybrid platforms used in quantum technologies. Examples include multimode cavity architectures \cite{zhang2016cavity,zhang2021nonreciprocal}, magnon-based quantum memories \cite{quantmem1}, and transducers linking microwave and optical photons \cite{transd1,transd2,transd3}. The ability to obtain analytic, physically transparent spectra—incorporating Floquet engineering, nonlinearities, and Zeeman shifts—provides a practical modeling tool for designing and optimizing quantum magnonic devices.

\section*{Acknowledgments}

TACF would  like to thank FAPESP (Process ID 2024/08133-4), CNPq (Process ID 141561/2023-8), and Campinas State University for funding. MCO acknowledges partial financial support from the National Institute of Science and Technology for Applied Quantum Computing through CNPq process No. 408884/2024-0 and  by FAPESP, through the  Center for Research and Innovation on Smart and Quantum Materials (CRISQuaM) process No. 2013/07276-1.

\appendix

\section{Calculations of the Perturbative Orders}
\label{pert_calc}

In this Appendix we provide the detailed Fourier–domain calculations used to obtain the first- and second-order perturbative terms appearing in the main text.  
All Fourier transforms are defined as
\[
\tilde{f}(\omega)=\mathcal{F}[f(t)] 
= \int_{-\infty}^{\infty} f(t)e^{i\omega t}\,dt,
\]
and convolution in frequency space is written as
\[
[A\star B](\omega)=\int A(\omega')\,B(\omega-\omega')\,d\omega'.
\]
Whenever the input fields appear, their Fourier transforms reduce to delta functions centered at $\pm\omega_{b,c}$, e.g.
\[
\xi_a^\pm(\omega) = \xi_a\, \delta(\omega \pm \omega_a).
\]

\section{First-Order Terms}

The first-order quantities involve convolutions of zeroth-order spectra.  
We evaluate each of them explicitly below.

\subsection{Term \texorpdfstring{$\mathcal{F}[\langle \hat{b}^{(0)}\hat{m}^{(0)}\rangle]$}{F[b0 m0]}}
\begin{align}
\mathcal{F}[\langle \hat{b}^{(0)}\hat{m}^{(0)}\rangle]
&= \mathcal{F}[\langle \hat{b}^{(0)}\rangle] \star
   \mathcal{F}[\langle \hat{m}^{(0)}\rangle] \nonumber\\
&= F_b^-(\omega)\xi_b^-(\omega)
   \star \bigl[iD^-(\omega)\xi_c^-(\omega)\bigr] \nonumber\\
&= i\xi_b F_b^-(\omega_b)\,
   D^-(\omega-\omega_b)\xi_c^-(\omega-\omega_b).
\end{align}

\subsection{Term \texorpdfstring{$\mathcal{F}[\langle \hat{b}^{(0)\dagger}\hat{m}^{(0)}\rangle]$}{F[b0† m0]}}
\begin{align}
\mathcal{F}[\langle \hat{b}^{(0)\dagger}\hat{m}^{(0)}\rangle]
&= F_b^{-*}(-\omega)\xi_b^*(-\omega)
   \star \bigl[iD^-(\omega)\xi_c^-(\omega)\bigr] \nonumber\\
&= i\xi_b^* F_b^{-*}(\omega_b)\,
   D^-(\omega+\omega_b)\xi_c^-(\omega+\omega_b).
\end{align}

\subsection{Term \texorpdfstring{$\mathcal{F}[\langle \hat{m}^{(0)\dagger}\hat{m}^{(0)}\rangle]$}{F[m0† m0]}}

\begin{align}
B_b^{(0)}(\omega)
&= \mathcal{F}[\langle \hat{m}^{(0)\dagger}\hat{m}^{(0)}\rangle] \nonumber\\
&= \bigl[-i D^{-*}(-\omega)\xi_c^*(-\omega)\bigr]
   \star \bigl[iD^-(\omega)\xi_c^-(\omega)\bigr]\nonumber\\
&= \xi_c^* D^{-*}(\omega_c)\,
   D^-(\omega+\omega_c)\xi_c^-(\omega+\omega_c).
\end{align}

\section{Second-Order Terms}

We next evaluate the terms contributing to the second-order magnon and cavity spectra.  
Each of them involves convolutions between first-order and zeroth-order contributions.

\subsection{Term \texorpdfstring{$B_m^{(1),-}(\omega)$}{Bm1-}}
This corresponds to $\mathcal{F}[\langle \hat{b}^{(1)}\hat{m}^{(0)}\rangle]$:
\begin{align}
B_m^{(1),-}(\omega)
&= \mathcal{F}[\langle \hat{b}^{(1)}\rangle]
   \star \mathcal{F}[\langle \hat{m}^{(0)}\rangle] \nonumber\\[3pt]
&= \bigl[-i\eta B_b^{(0)}(\omega)F_b^-(\omega)\bigr]
   \star \bigl[iD^-(\omega)\xi_c^-(\omega)\bigr] \nonumber\\[3pt]
&= \eta |\xi_c|^2 |D^-(\omega_c)|^2\,
   F_b^-(\omega-\omega_c)\,D^-(\omega)\,\xi_c^-(\omega).
\end{align}

\subsection{Term \texorpdfstring{$B_m^{(1),+}(\omega)$}{Bm1+}}
\begin{align}
B_m^{(1),+}(\omega)
&= \mathcal{F}[\langle \hat{b}^{(1)\dagger}\hat{m}^{(0)}\rangle] \nonumber\\
&= \eta \xi_c^2 D^-(\omega_c)^2
   F_b^*(\omega_c-\omega)\,
   D^{-*}(2\omega_c-\omega)\,
   \xi_c^-(\omega).
\end{align}

\subsection{Term \texorpdfstring{$\mathcal{F}[\langle \hat{b}^{(0)}\hat{m}^{(1)}\rangle]$}{F[b0 m1]}}
\begin{align}
\mathcal{F}[\langle \hat{b}^{(0)}\hat{m}^{(1)}\rangle]
=& F_b^-(\omega)\xi_b^-(\omega)
   \star \bigl[\eta D^-(\omega) B_m^{(0)}(\omega)l_c^-(\omega)\bigr]\nonumber\\
=& i\eta|\xi_b|^2|F_b^-(\omega_b)|^2\,
   D^-(\omega-\omega_b)\,
   l_c^-(\omega-\omega_b)\nonumber\\
  & \times
   D^-(\omega)\xi_c^-(\omega).
\end{align}

\subsection{Term \texorpdfstring{$\mathcal{F}[\langle \hat{b}^{(0)\dagger}\hat{m}^{(1)}\rangle]$}{F[b0† m1]}}
\begin{align}
\mathcal{F}[\langle \hat{b}^{(0)\dagger}\hat{m}^{(1)}\rangle]
=& F_b^+(\omega)\xi_b^+(\omega)
   \star \bigl[\eta D^-(\omega)B_m^{(0)}(\omega)l_c^-(\omega)\bigr]\nonumber\\
=& i\eta|\xi_b|^2|F_b^-(\omega_b)|^2\,
   D^-(\omega+\omega_b)\,
   l_c^-(\omega+\omega_b)\nonumber\\
   &\times
   D^-(\omega)\xi_c^-(\omega).
\end{align}
\bigskip

\noindent
The expressions above determine all quantities entering the second-order reflectance formula in the main text.
They also reveal the structure of the Floquet sidebands:  
each sideband frequency arises from the convolution of a cavity term at $\omega$ with a modulation or magnon term at $\omega\pm\omega_b$, which is why the perturbative expansion captures the experimentally observed sideband pattern.

\section{Derivation of the Re-normalized Magnon Frequency via a Coherent Modulation Field\label{app_D}}

If we take the modulation field to be a coherent drive, the ground-state Zeeman term becomes a time-dependent contribution plus a fluctuation term
\begin{align}
    \hat{U}_b=-g_bNs(\hat{b}+\hat{b}^\dagger)\rightarrow -&\dfrac{1}{2}G_b \epsilon^{-1}(\beta+\beta^*)\mathbb{I}\\-&\dfrac{1}{2}G_b\epsilon^{-1}(\delta\hat{b}+\delta \hat{b}^\dagger).
\end{align}Ignoring the latter term, we find that the full equation of motion of $\beta$ is
\begin{align}
    \dot{\beta}=-\left(\dfrac{\Gamma_b}{2}+i\omega_b\right)\beta+\sqrt{\kappa_b}\beta_{\text{in}}e^{-i\omega_bt} +\dfrac{i}{2}G_b\epsilon^{-1}.
\end{align}where  $\beta_{\text{in}}=|\beta_{\text{in}}|e^{-i\phi_b}$. Using the change of variables $\beta'=\beta e^{(\Gamma_b/2+i\omega_b)t}$, we simplify this equation to
\begin{align}
    \dot{\beta}' =\sqrt{\kappa_b}\beta_{\text{in}}e^{
    \Gamma_bt/2}+\dfrac{i}{2}G_b\epsilon^{-1}e^{(\Gamma_b/2+i\omega_b)t},
\end{align}
whose solution, re-expressed in terms of $\beta$, is
\begin{align}
    \beta(t)=\dfrac{2\beta_{\text{in}}}{\sqrt{\kappa_b}}e^{-i\omega_bt}+\dfrac{iG_b\epsilon^{-1}}{\Gamma_b+2i\omega_b}\left(1-e^{-(\Gamma_b/2+i\omega_b)t}\right).
\end{align}
The stationary solution is, therefore,
\begin{align}
     \beta_{ss}=\beta(t\rightarrow \infty)=&\dfrac{2\beta_{\text{in}}}{\sqrt{\kappa_b}}e^{-i\omega_bt}+\dfrac{iG_b\epsilon^{-1}}{\Gamma_b+2i\omega_b}\\
     =&\dfrac{2\beta_{\text{in}}}{\sqrt{\kappa_b}}e^{-i\omega_bt}+G_b\epsilon^{-1}\left(\dfrac{i\Gamma_b+2w_b}{\Gamma_b^2+4\omega_b^2}\right).
\end{align}
This means that the Zeeman Interaction induces a static back-action onto the field which in turn renormalizes the natural magnon frequency,
\begin{align}
    \hat{U}_{mb}=&\epsilon G_b(\hat{b}+\hat{b}^{\dagger})\hat{m}^{\dagger}\hat{m}\\
    \rightarrow &\epsilon \dfrac{2|\beta_{\text{in}}|}{\sqrt{\kappa_b}}\cos(\omega_bt+\phi_b)\hat{m}^{\dagger}\hat{m}+\dfrac{4G_b^2\omega_b}{\Gamma_b^2+4\omega_b^2}\hat{m}^{\dagger}\hat{m}\\
    &=\epsilon \dfrac{2|\beta_{\text{in}}|}{\sqrt{\kappa_b}}\cos(\omega_bt+\phi_b)\hat{m}^{\dagger}\hat{m}+\Delta_b\hat{m}^{\dagger}\hat{m}.
\end{align}
From the definition of the static Zeeman coupling $g_0=\gamma B_0$, we can define $g_{\text{eff}}=\gamma B_{\text{eff}}=\gamma\left(B_0+\dfrac{\Delta_b}{\gamma}\right)$  such that this shift acts as an effective field re-normalization.

\bibliography{apssamp}% Produces the bibliography via BibTeX.

\end{document}